\documentclass[twocolumn,prl,aps,superscriptaddress,showpacs]{revtex4}

\usepackage{dcolumn}
\usepackage{amsmath}
\usepackage{epsfig}
\newlength{\figwidth}
\newlength{\figwidthb}
\begin{document}


\title{Charge-transfer exciton in La$_2$CuO$_4$ probed with resonant inelastic x-ray scattering}
\author{D. S. Ellis}
\affiliation{Department of Physics, University of Toronto, Toronto,
Ontario M5S~1A7, Canada}
\author{J. P. Hill}
\affiliation{Department of Physics, Brookhaven National Laboratory,
Upton, New York 11973}
\author{S. Wakimoto}
\affiliation{Advanced Science Research
   Center, Japan Atomic Energy Research Institute, Tokai, Ibaraki
   319-1195, Japan}
\author{R. J. Birgeneau}
\affiliation{Department of Physics, University of California,
Berkeley, California 94720-7300}
\author{D. Casa}
\affiliation{CMC-XOR, Advanced Photon Source, Argonne National
Laboratory, Argonne, Illinois 60439}
\author{T. Gog}
\affiliation{CMC-XOR, Advanced Photon Source, Argonne National
Laboratory, Argonne, Illinois 60439}
\author{Young-June Kim}
\email{yjkim@physics.utoronto.ca}
\affiliation{Department of
Physics, University of Toronto, Toronto, Ontario M5S~1A7, Canada}

\date{\today}

\begin{abstract}
We report a high-resolution resonant inelastic x-ray scattering
study of $\rm La_2CuO_4$.  A number of spectral features are
identified that were not clearly visible in earlier lower-resolution
data. The momentum dependence of the spectral weight and the
dispersion of the lowest energy excitation across the insulating gap
have been measured in detail.  The temperature dependence of the
spectral features was also examined. The observed charge transfer
edge shift, along with the low dispersion of the first charge
transfer excitation are attributed to the lattice motion being
coupled to the electronic system. In addition, we observe a
dispersionless feature at 1.8 eV, which is associated with a d-d
crystal field excitation.
\end{abstract}


\maketitle


As the prototypical parent insulating compound, $\rm La_2CuO_4$ has
drawn much attention as a starting point for the study of
high-temperature superconductors.  $\rm La_2CuO_4$ is well described
as an antiferromagnetic insulator with a $d_{x^2-y^2}$ hole
localized at the Cu site due to the strong electron correlations. In
particular, this material is classified as a charge-transfer (CT)
insulator, since the lowest energy charge excitation across the
insulating gap corresponds to transferring an electron from oxygen
to a neighboring copper.  This CT excitation creates a Cu$^{1+}$ ion
and an oxygen 2p hole that forms a Zhang-Rice singlet (ZRS)with the
neighboring copper spin, the two copper sites forming a bound state
\cite{Zhang88}. Over the years, exciton formation of these
electron-hole pairs has been considered by several authors as a
possible low energy elementary excitation in this half-filled copper
oxide plane
\cite{Clarke93,Simon96,Zhang98,Hanamura00,Wrobel02,Moskvin02,Moskvin05}.
The strong magnetic interaction in this system has been pointed to
as the origin of the exciton binding energy \cite{Clarke93} as well
as the large dispersion of the excitation \cite{Zhang98}. Indeed the
apparently large dispersion observed in $\rm Sr_2CuO_2Cl_2$ using
electron energy loss spectroscopy (EELS) was attributed to a CT
exciton with a small effective mass \cite{Wang96}.

  Experimentally, optical spectroscopy and Raman scattering studies
have been carried out to address the nature of the excitation
spectrum near this CT gap \cite{Ohana89, Falck92, Lovenich01,
Salamon95}. Ohana et al. directly observed the coupling of the
electronic excitations to the lattice system in the resonance of
phonon Raman lines at laser energies around the CT energy (2.14 eV)
and a low-energy shoulder at 1.9 eV \cite{Ohana89}. Falck et al.
found that a polaron model was able to reproduce their reflectivity
data very well for a wide temperature range \cite{Falck92}. However,
despite extensive experimental and theoretical studies to date, the
nature of the CT exciton is still controversial. Outstanding
questions include whether the electron-hole pair actually form a
bound exciton state, or if they remain as a resonance state within
the particle-hole continuum. Another question relates to the
relative role played by magnetic interactions and/or phonons in the
exciton formation and dispersion. For each of these questions,
valuable information can be gained by studying the momentum
dependence of the charge transfer excitations. In recent years,
resonant inelastic x-ray scattering (RIXS) has drawn considerable
interest as a valuable spectroscopic tool for studying the
momentum-dependence of various electronic excitations in insulating
cuprates such as these excitons
\cite{Hill98,Abbamonte99,Hasan00,Kotani01,Kim02,Lu05,Lu06,Collart06}.
However, these earlier studies have been hampered by relatively poor
instrumental energy resolution.

In this Letter, we report detailed studies of electronic excitations
near the CT gap in the prototypical cuprate compound, $\rm
La_2CuO_4$, using the highest energy resolution to date for RIXS,
130 meV (full-width at half-maximum).  We observe several new
features which were previously not clearly visible.  We find that 1)
the RIXS spectra are composed of multiple spectral features which
exhibit complex dispersion and spectral weight change as a function
of momentum transfer; 2) the two lowest energy features observed
correspond to a dispersionless $dd$-excitation at 1.8 eV and a CT
exciton at 2.2 eV; 3) the latter disperses with a bandwidth of $\sim
0.3$ eV in the ($\pi$,0) direction and of at least that much along
($\pi$,$\pi$) where its intensity decreases rapidly as one goes from
the zone centre towards ($\pi$,$\pi$) and 4) the edge of the CT
excitation shifts to lower energy for increased temperatures in
accordance with previous optical studies \cite{Falck92}. We argue
that these results are consistent with a picture of a bound exciton
state in the presence of strong \textit{e-ph} coupling.

  The RIXS experiments were carried out at the Advanced Photon Source on the
undulator beamline 9IDB.  A detailed description of the experimental
setup was reported in Ref.~\cite{Hill05}.  The floating-zone grown
$\rm La_2CuO_4$ sample was cooled to 27~K by a closed cycle
refrigerator to reduce the phonon contribution to the background. We
used the same experimental geometry as that used in
Ref.~\cite{Kim02}.

\begin{figure}
\centering
\epsfig{file=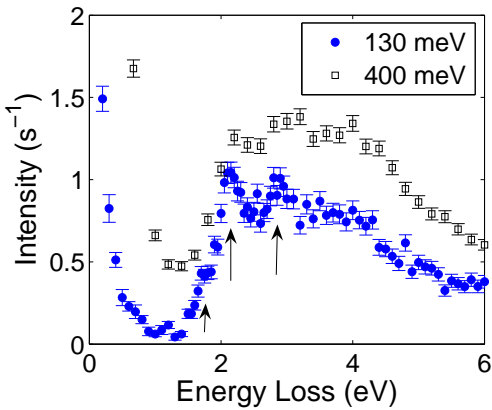,width=\figwidth}
\caption{(Color online) Comparison of RIXS spectra obtained using
400 meV resolution (squares, from Ref.~\cite{Kim02}) and 130 meV
resolution (circles) used in the present work.  The 400 meV spectrum
has been scaled by a factor of $1/5$ to display comparable inelastic
intensity with the 130 meV spectra.}  \label{fig:resolution}
\end{figure}

The improvement in energy resolution is clearly visible in
Fig.~\ref{fig:resolution}, in which we compare the RIXS spectra of
$\rm La_2CuO_4$\ obtained with two different resolutions. The open
square symbols represent data taken from Ref.~\cite{Kim02}, obtained
with 400 meV resolution, while the filled circles represent the new
data obtained with 130 meV resolution.  Both spectra were measured
with the same incident energy ($E_i=8992$ eV) \cite{energynote} and
momentum transfer {{\bf Q}=(3 0 0)} in tetragonal notation. Note
that the tail of the elastic line is drastically reduced in the
energy range below 2 eV.  Although the overall shapes of the
spectral features are almost identical, the improved energy
resolution allows us to see several sharp features in the spectrum.
Specifically, one can identify a sharp peak at the energy
corresponding to the CT gap of this material at 2.2 eV, and several
spectral features at higher energies, including a prominent peak at
2.9 eV.  These features have been observed in a recent study by Lu
et. al. \cite{Lu06}, where the incident energy was varied to
elucidate the different peaks. In addition, a low-energy shoulder of
the main peak is, for the first time at q=0, resolved at around 1.8
eV (this feature has been recently observed near the zone boundary
\cite{Collart06}). Since the overall incident energy dependence of
the spectral features at zero momentum transfer was found to be
consistent with earlier studies \cite{Kim02}, we focus our attention
on the momentum dependence at fixed incident energy in the
following.

\begin{figure}
\centering \epsfig{file=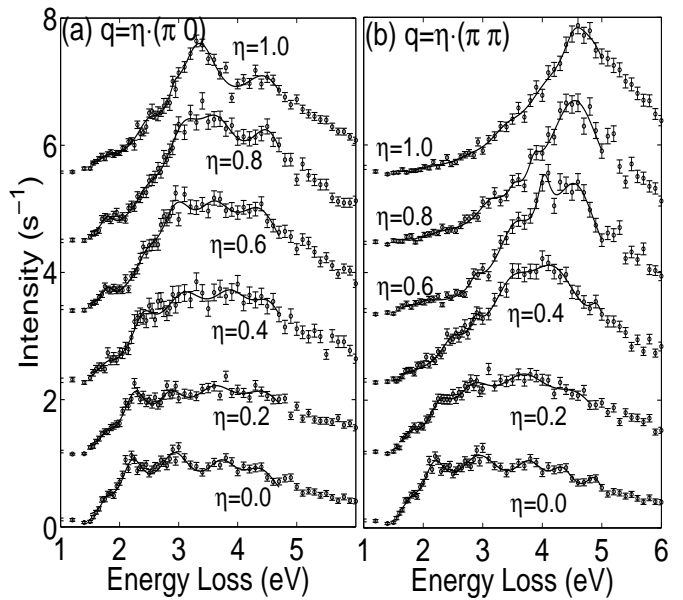,width=3.9in, height=3.1in }
\caption{Momentum dependence of the RIXS spectrum along the (a)
($\pi$, 0) and (b) ($\pi$, $\pi$) directions.  The incident energy
is 8992.5 eV.  All curves are normalized by the elastic peak
intensity during each scan, and shifted along the intensity axis for
clarity.} \label{fig:momentum}
\end{figure}

To investigate the momentum dependence, energy loss scans were taken
with the fixed incident energy at various reciprocal space points in
the two high-symmetry directions: ($\pi$,0) along the Cu-O bond, and
($\pi$,$\pi$) at a 45 degree angle.  Each scan was normalized by the
intensity of the elastic peak.  Figure~\ref{fig:momentum} shows the
development of the spectra as the momentum transfer {\bf q}
increases away from zone center. The zone boundary scans at
($\pi$,0) and ($\pi$,$\pi$) look similar to earlier low-resolution
data; that is, there exist two well defined peaks at ($\pi$,0),
while only one prominent peak appears at ($\pi$,$\pi$).  The
progression of the spectral features as a function of momentum
transfer involves both actual shifts of the peak positions as well
as intensity changes in the respective peaks. In order to analyze
the dispersion quantitatively, the spectra were fit to multiple
Lorentzian peaks (solid lines in Fig.~\ref{fig:momentum}).

\begin{figure}
\centering \epsfig{file=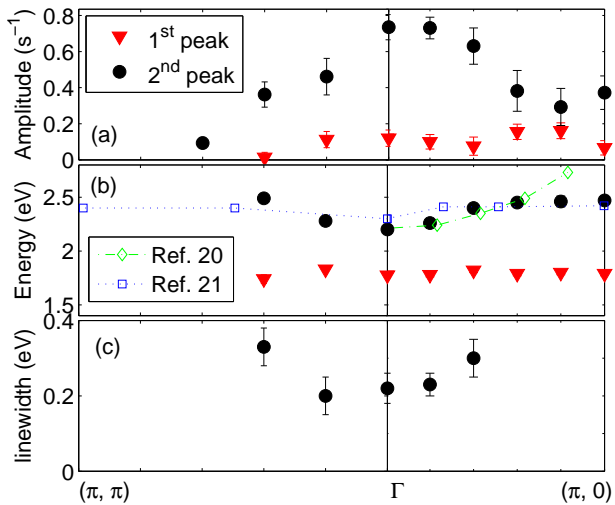,width=\figwidth}
\caption{(Color online) Momentum dependence of fitting parameters:
(a) Spectral weights, (b) center positions, and (c) Lorentzian
widths (FWHM). Due to the vanishing spectral weights, the energies
and widths were only reliably determined until about half way across
the Brillouin zone in the ($\pi$, $\pi$) direction.  Included in the
energy dispersion plot is a comparison with the results from other
recent studies \cite{Collart06, Lu06}} \label{fig:dispersion}
\end{figure}

The results of the fits for the two lowest energy features are shown
in Fig.~\ref{fig:dispersion}.  The salient feature of the 1.8 eV
peak is its lack of energy dispersion, which suggests that this
excitation is spatially localized.  The most likely candidate for
this excitation is a crystalline-field $dd$ excitation.  We note
that charge transfer with the surrounding $O 2p$ states of the same
symmetry would also be involved \cite{Moskvin05, Salamon95}. This
peak assignment is consistent with earlier studies using different
experimental techniques, such as large-shift Raman scattering
\cite{Salamon95}, Cu M-edge RIXS \cite{Kuiper98}, and L-edge RIXS
\cite{Ghiringhelli04}. In addition, the 1.8 eV shoulder was also
observed near the zone boundary in a recent K-edge RIXS study
\cite{Collart06}, and likewise attributed to a crystal field
excitation.  It is interesting to note that the $dd$-excitation is
much stronger than the CT excitation (at 2.2 eV) in the L-edge
spectra, and vice versa in our K-edge spectra. Since the
intermediate state of the L-edge RIXS directly involves Cu
$d$-orbitals, it would have large overlap with the final state of
the $dd$-excitation, and generate large $dd$ intensity as predicted
by Tanaka et al. \cite{Tanaka93}.

The next peak, corresponding to the main sharp feature at $\omega
\sim 2.2 $ eV in Fig.~\ref{fig:resolution}, was seen in previous
RIXS and optical conductivity studies and widely associated with the
excitation across the CT gap \cite{Uchida91}. The dispersion of the
CT excitation is shown in Fig.~\ref{fig:dispersion}(b).  A direct
gap is exhibited, with a total bandwidth along the ($\pi$,0)
direction of around 0.3 eV. This observation of a small bandwidth is
in contrast to the earlier report of $\sim 1$ eV bandwidth observed
with lower-resolution setup \cite{Kim02}, but is consistent with the
small bandwidth ($\leq$ 0.5 eV) observed in more recent RIXS studies
\cite{Lu05,Lu06,Collart06} as shown in Fig.
~\ref{fig:dispersion}(b).  We note that our energies near ($\pi$,0)
seem to match well with those of Ref. \cite{Lu06}.  In contrast, the
dispersion we observe matches very well with Ref. \cite{Collart06}
from low q's up to two thirds of the way towards the zone boundary,
but then the two studies disagree for the last point of Ref.
\cite{Collart06}.  This is not entirely surprising, since the 2.2 eV
peak loses intensity near the zone boundary and becomes a shoulder
of the higher energy peaks, making energy determination more
uncertain, especially in the case of lower resolution.

  The observed bandwidth is in fact very close to that of the ZRS band
of the insulating cuprates \cite{Wells95}. Along the ($\pi$,$\pi$)
direction, the intensity of the 2.2 eV peak decreases, and the peaks
eventually becomes unobservable around the halfway point across the
zone. Similar dramatic momentum dependencies of the spectral weight
have been predicted for certain CT excitons \cite{Zhang98,
Moskvin05}. We also note that the peak width is not limited by
resolution, and broadens as q increases, as shown in Fig.
~\ref{fig:dispersion}(c), suggesting that the excitation becomes
less well-defined at larger momentum transfers. Such a result would
be expected if there are more decay channels available at higher q.

\begin{figure}
\centering \epsfig{file=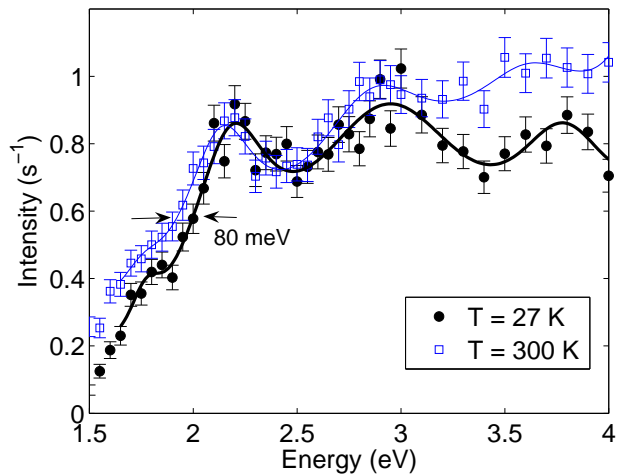,width=\figwidth}
\caption{(Color online) Comparison of the zone center spectra at T =
27~K and T = 300~K.  The two spectra were scaled to match their
intensities at 10 eV energy loss.} \label{fig:temperature}
\end{figure}

We have also obtained the zone center spectrum at 300 K which is
compared to the 27~K scan in Fig.~\ref{fig:temperature}. Since the
elastic intensities vary at different temperatures, the two spectra
were in this case scaled to match the intensities of their Cu $\rm K
\beta_5$ emission lines, which are at 10 eV.  Note that the CT peak
intensity also happens to match when this normalization scheme is
used, which is appropriate for comparing shifts in the CT gap edge.
A red shift of $\sim80$ meV is seen in the CT gap edge, measured at
$\omega$=2 eV as shown in the figure. Sources of error in this shift
determination include uncertainty in the energy loss scale, and
possible incident energy variations between the two scans.  The
effect of the latter was found to be negligible; even for relatively
large changes in the incident energy (500 meV), the fitted edge of
the CT peak did not shift by more than 5 meV.  One contribution to
the error in energy loss scale between the two scans is the
zero-loss reference, determined by fitting the elastic line, and is
accurate to within 10 meV.  Thus, we conclude that the observed
shift is mostly due to the temperature change, and agrees with the
$\sim 100$ meV shift observed in optical reflectivity measurements
\cite{Falck92}. We would like to point out that although the RIXS
spectrum has often been compared with the corresponding optical
spectrum,  it has been difficult to draw direct correlation between
the features observed in the two spectroscopies. The same thermal
behavior of the 2.2 eV peak from the RIXS and the reflectivity
measurement \cite{Falck92} implies that these peaks have the same
origin.

One of the outstanding questions is whether or not the 2.2 eV peak
seen in these data indeed corresponds to a two-copper-site, bound
exciton as described by a Zhang-Ng-type model \cite{Zhang98} . One
might use the non-zero peak width, indicative of a relatively short
lifetime, to argue against a bound exciton-type picture. However we
argue that in fact a bound exciton is a good description of the
data.  As possible explanations for the finite peak width, we point
to the possibility of multiple excitations within the instrumental
resolution, or alternatively, intrinsic broadening mechanisms
associated with the antiferromagnetic lattice \cite{Doniach71}.  A
third possibility is that the observed broadening of the RIXS peak
as q is increased could result from the unbinding of the exciton. In
this proposed scenario, the exciton energy lies just below the upper
Hubbard band. Increasing q causes the bound exciton to disperse into
the electron-hole continuum, decreasing the lifetime of the exciton
and broadening the associated peak. This picture is somewhat similar
to the situation in the 1D system $\rm Sr_2CuO_3$ (Fig.~2(b)
Ref.\cite{Neudert98}), but the bound exciton exists near the zone
center in $\rm La_2CuO_4$.  Note that the binding energy of this
exciton must be very small, since a large increase of the
photoconductivity was observed just above the gap \cite{Thio90}.
With these possible explanations for the observed non-zero peak
width, it is reasonable to suppose that the 2.2 eV peak is indeed
the lowest-energy CT exciton.

The relatively small overall magnitude of the observed dispersion
would be expected if the excitations are strongly coupled to
phonons, which will give them a high effective mass. On the other
hand, a purely electronic model with only magnetic degrees of
freedom (such as that of Ref.\cite{Zhang98}) predicts much larger
dispersion than that observed, since singlet exciton propagation
does not disturb a magnetically ordered spin background. Thus, our
results support a picture in which the \textit{e-ph} interaction
plays an important role in the exciton dispersion.  This is also
consistent with the optical reflectivity study \cite{Falck92}.
Presumably the same mechanism, namely a temperature dependent
\textit{e-ph} interaction is also responsible for the observed
red-shift of the RIXS peak \cite{Fan51,Frohlich50}.

Finally, the higher energy excitations are somewhat more difficult
to interpret, given the large overlap with each other and
uncertainties in the fitting, and we can provide at most
speculations about them. However, there is a clear peak in
Fig.~\ref{fig:momentum} at 2.9 eV which shows little or no energy
dispersion, and whose amplitude decreases along the ($\pi$,$\pi$)
direction but not along ($\pi$,0), where it can be seen as a
shoulder to the main peak at zone boundary. This could be the
dispersionless $b_1$$_g$$e_u$($\pi$) 2-center exciton discussed by
Moskvin et. al. \cite{Moskvin05}.  At the zone boundaries, there are
strong peaks at 4.5 eV for q=($\pi$,$\pi$) and 3.3 eV for
q=($\pi$,0). The narrowing of the latter feature gives the
appearance of neighboring modes drawing together as q approaches the
zone boundary, likely involving strong interaction between the modes
\cite{Moskvin05}.

In summary, we have studied the electronic excitations near the
charge transfer gap of $\rm La_2CuO_4$ using high resolution
resonant inelastic x-ray scattering, observing a dispersionless $dd$
excitation at 1.8 eV, and a weakly dispersive CT exciton at 2.2 eV
at the Brillouin zone center, as well as higher energy peaks. Away
from the zone center, the CT exciton peak broadens and disperses to
higher energy, while losing its spectral weight. Its observed
bandwidth is similar to that of the ZRS \cite{Wells95}, and is much
smaller than previously reported values, but in accordance with
recent studies \cite{Collart06,Lu06}. This small bandwidth of the
exciton dispersion, as well as the observed temperature dependence,
underscores the importance of the \textit{e-ph} interaction in
insulating cuprates. It is clear that further theoretical study of
exciton behavior in the presence of strong \textit{e-ph}
interactions would be useful for quantitative understanding of the
electron correlations in cuprates.

We would like to thank K. H. Ahn, C. Kim, S. Larochelle, G.
Sawatzky, K. Tsutsui, T. Tohyama, J. van den Brink, and M. van
Veenendaal for invaluable discussions. The work at University of
Toronto was supported by Natural Sciences and Engineering Research
Council of Canada. The work at Brookhaven was supported by the U. S.
DOE, Office of Science Contract No. DE-AC02-98CH10886. Use of the
Advanced Photon Source was supported by the U. S. DOE, Office of
Science, Office of Basic Energy Sciences, under Contract No.
W-31-109-ENG-38.  R.J.B. is supported at Lawrence Berkeley
Laboratory by the Office of Basic Energy Sciences, U.S. DOE under
Contract No. DE-AC03-76SF00098.



\end{document}